\documentclass[%
 aip,
 apl,
 groupedaddress,
 amsmath,amssymb,
 reprint,%
]{revtex4-1}

\usepackage{graphicx}
\usepackage{dcolumn}
\usepackage{bm}
\usepackage{braket}
\usepackage[utf8]{inputenc}
\usepackage[T1]{fontenc}
\usepackage{mathptmx}
\usepackage{etoolbox}
\usepackage{xcolor}
\usepackage{siunitx}
\DeclareSIUnit{\dBm}{dBm}
\usepackage[hidelinks]{hyperref}

\newcommand{\mc}[1]{{\textcolor{black}{#1}}}
\newcommand{\so}[1]{{\textcolor{black}{#1}}}
\newcommand{\jso}[1]{{\textcolor{black}{#1}}}

\makeatletter
\def\emailoutput{{\ensuremath{{}^\text{a)}}}Authors to whom correspondence should be addressed:}
\def\emailseparation{ }
\patchcmd{\titleblock@produce}
  {\frontmatter@RRAPformat}
  {\frontmatter@RRAPformat{\produce@RRAP{\emailoutput}} \frontmatter@RRAPformat}
  {}{}

\def\@fixname#1#2#3{%
    \def\@author{{#1#3}{#2}}
}
\def\@email#1#2#3{%
 \endgroup
 \expandafter\@fixname\@author{#3}
 \appto {\emailoutput}{\emailseparation\href{mailto:#2}{#2}\patchcmd{\emailseparation}{,}{,}{}{\preto{\emailseparation}{,}}}{}
 
}
\makeatother

\begin{document}

\preprint{AIP/123-QED}

\title[Sample title]{\jso{Distant RF field sensing with a passive Rydberg-atomic transducer  }}
\author{J. Susanne Otto}
\email{susi.otto@otago.ac.nz}{\ensuremath{{}^\text{a)}}}
\author{Matthew Chilcott}%
\author{Amita B. Deb}%

\author{Niels Kj{\ae}rgaard}%
 \email{niels.kjaergaard@otago.ac.nz}{\ensuremath{{}^\text{a)}}}
\affiliation{%
Department of Physics, QSO—Quantum Science Otago, and Dodd-Walls Centre for Photonic and Quantum Technologies,
University of Otago, Dunedin, New Zealand
}%

\date{\today}

\begin{abstract}
We combine a rubidium vapor cell with a corner-cube prism reflector to form a passive RF \jso{transducer}, allowing the detection of microwave signals at a location distant from the active components required for atomic sensing. This compact \jso{transducer} has no electrical components and is optically linked to \jso{an} active base station by a pair of free-space laser beams that establish an electromagnetically induced transparency scenario. Microwave signals at the \jso{transducer} location are imprinted onto an optical signal which is detected at the base station. Our \jso{sensing} architecture \mc{with a remote standalone transducer unit} adds important flexibility to Rydberg-atom based sensing technologies, which are currently subject to significant attention. We demonstrate a \jso{\SI{\sim 30}{\metre}} link with no particular effort and foresee significant future prospects of achieving a much larger separation between \so{the} \jso{transducer and the base station}.
\end{abstract}

\maketitle

In 2012, a seminal work\cite{Sedlacek2012} by Sedlacek \textit{et al.} demonstrated the use of Rydberg-excited $\rm^{87}Rb$ atoms in a glass vapor cell as a sensitive detector for microwave fields. In their scheme, the presence and strength of microwave radiation resonant with the transition between two Rydberg states was measured by sending counter-propagating blue and infrared laser beams through the vapor cell. In the absence of microwave radiation, the blue coupling light would establish electromagnetically induced transparency (EIT) for the infrared probe light, while an incoming microwave field of constant amplitude would split the EIT transmission peak into two, and from the peak separation of these two Autler-Townes (AT) peaks, the microwave field strength could be inferred. Because atomic transitions lie at the heart of this ingeniously simple experimental scheme, it offers \so{SI} traceable and calibration free electrometry\jso{~\cite{Holloway2017a}.}

The decade following the original work of Ref.~\onlinecite{Sedlacek2012} witnessed a number of developments.
For example, the sensitivity of the on-resonant EIT/AT technique can be greatly improved using a superheterodyne receiver architecture \cite{Jing2020} and an optical ground-state repumping technique\cite{Prajapati2021}, making it possible to detect fields down to \SI{5}{\micro\volt\per\meter}.  
In the other extreme, at large RF field strengths the AT splitting is no longer linear with the strength of the applied field due to the AC Stark shift, which scales as the field strength squared.
Instead, this shift can be measured to infer the field strengths of sufficiently large fields \cite{Anderson2016,Paradis2019}.
Measurements of the AC Stark shift also allow detection of frequencies outside the discrete set of atomic Rydberg transitions \cite{Hu2022} extending the technique to a continuous RF spectrum. Rydberg-atomic systems for communications \cite{Meyer2018,Cox2018,Song2018,Deb2018} constitutes a  particular application that leverages the technique of Ref.~\onlinecite{Sedlacek2012} and fundamental working principles of analog and digital communication have been demonstrated in a range of atom-based systems \cite{Song2019,Holloway2019c,Holloway2020,Anderson2020}.
\begin{figure*}
\includegraphics[width=0.99\textwidth]{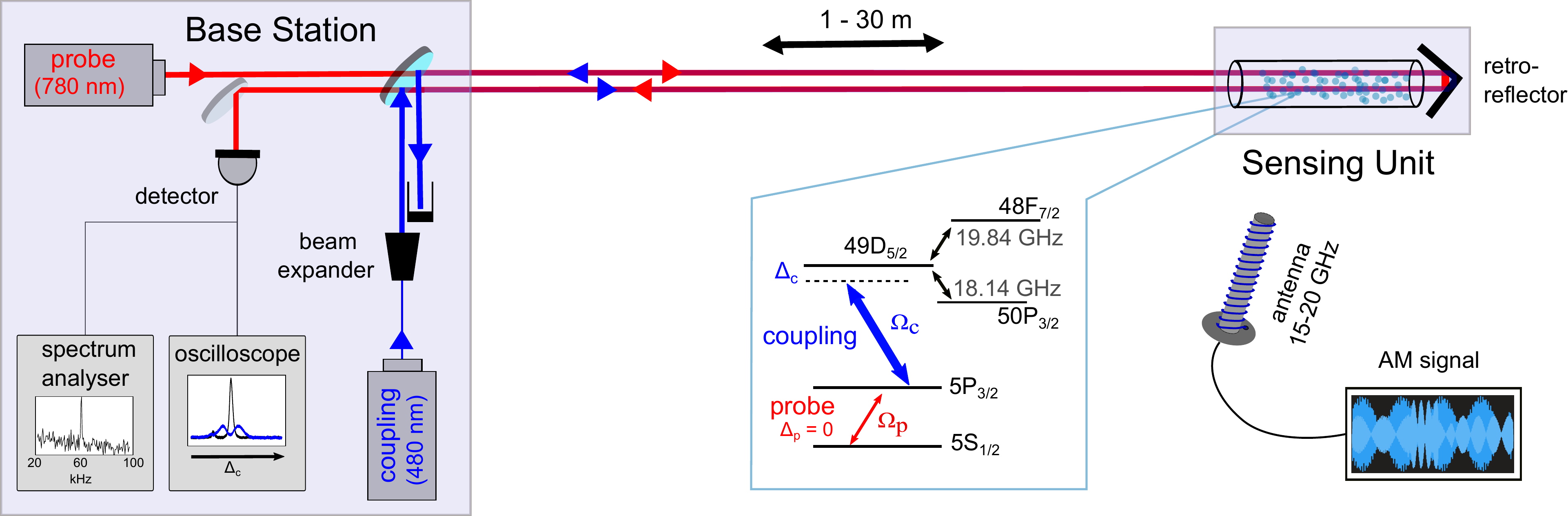}
\caption{\label{fig:setup}Schematic of the experimental setup. For clarity, the laser driver electronics and the setup for frequency stabilizing the lasers are not included in the drawing of the base station. The inset shows the relevant atomic transitions in $^{87}$Rb, where \jso{$\Omega_\text{c}$, $\Omega_\text{p}$ and $\Delta_\text{c}, \Delta_\text{p}$ are the Rabi frequencies and detunings} of the coupling and probe fields, respectively. }
\end{figure*}

Although the systems mentioned above allow for a large variety of RF measurements and applications, they are almost invariably confined to an optical table in a laboratory due to the requirement to counter-align the two (or more) lasers within the atomic vapor cell. Early attempts of overcoming this constraint and separating the laser generation and detection from the RF probe---the vapor cell---made use of optical fibers bonded to the glass cell \cite{Holloway2017,Simons2018,Mao2023}. These fibers carried the light fields to the vapor cell, and guided back the probe field transmitted through the atomic vapor which carried the information. However, such a system offers limited portability, \jso{due to the} physical connections of the optical fibers to the vapor cell. Moreover, \jso{losses in the fibers attenuate the optical fields} and for high coupling efficiencies \jso{into the return fiber}, the probe beam needs to be strongly focused which increases transit-time broadening of the signal. In this \jso{Letter}, we \mc{present} a proof-of-concept setup capable of \jso{increased} mobility, replacing the fiber access to the vapor cell with two free-space laser beams and a corner-cube prism reflector, that reflects the probe beam back to a photodetector. Without any significant effort, we can \jso{deploy} our portable atomic RF probe to \jso{sense} fields at a distance of \jso{exceeding \SI{30}{\meter}} from the active components---the lasers and \jso{photodetector}. 
\newline\newline
Figure~\ref{fig:setup} shows a schematic of the experimental setup, \jso{highlighting the base station, and the separate portable transducer, in the following referred to as ``sensing unit''. Also shown are the relevant atomic levels of $^{87}$Rb involved in the RF-to-optical transduction.} The base station contains all the required elements to prepare two light fields for a two-photon Rydberg EIT system, i.e., \jso{a coupling and probe laser}, their driver electronics, and a setup for the frequency stabilization of the \jso{two} lasers. 
Specifically, we employ a coupling laser with a wavelength of \SI{480}{\nano\metre} and a maximum power of \jso{\SI{10}{\milli\watt}}. After passing through a beam expander, the coupling beam has a $1/e^2$ beam diameter of \SI{\sim 6}{\milli\metre} corresponding to a Rayleigh length of about \SI{60}{\metre}. \jso{For parts of our demonstration}, the coupling laser is stabilized to a high-finesse cavity using the Pound-Drever-Hall technique, which results in a root-mean square (RMS) linewidth below \SI{100}{\kilo\hertz}. \so{Alternatively}, the coupling laser could be locked to the EIT line of a separate reference cell, \jso{or, for a more compact design of the base station the self-locking technique described in Ref.~\onlinecite{Fancher2023} could be employed}. The probe laser at \SI{780}{\nano\metre} is stabilized to the $^{87}$Rb D$_2$-line 5S$_{1/2} (\text{F}=2) \leftrightarrow \text{5P}_{3/2}$(F'=3) via saturated absorption spectroscopy, resulting in a RMS laser linewidth of \SI{\sim 300}{\kilo\hertz}. \jso{At the base station, the probe has a $1/e^2$ beam diameter of \jso{\SI{2.5}{\milli\metre}} corresponding to a Rayleigh length of about \SI{6}{\metre}}. A dichroic mirror combines the linearly polarized probe and coupling beam which are leaving the base station parallel to one another. The base station also hosts all equipment needed to detect and analyze the returning probe light field: \jso{a photodetector, a spectrum analyzer and an oscilloscope. }

The stand-alone \jso{sensing unit} is optically linked to the base station via the two free space laser beams and only incorporates two elements: a \SI{150}{\milli\metre}-long \jso{and \SI{27}{\milli\metre}-wide} cylindrical vapor cell containing a rubidium vapor at room temperature and a corner-cube prism reflector\footnote{Thorlabs PS975-A}. The prism reflector allows us to accommodate two pairs of counter-propagating probe and coupling beams, passing through the vapor cell with minimal alignment effort. We note that such a setup benefits from a large sensing volume created by two detection areas in the vapor cell, as discussed in Ref.~\onlinecite{Otto2021a}, and that no lenses are needed for our \jso{sensing unit}. When \jso{coupled} to a Rydberg state via the \jso{optical} two-photon transition, the rubidium atoms become sensitive to RF \jso{radiation. In particular, this happens for RF frequencies that couple the Rydberg level resonantly to other nearby Rydberg levels}. For sufficiently large fields, the atoms are receptive to a broad range of RF signals via the AC Stark shift. In our demonstration, a home-built helical antenna broadcasts an amplitude modulated (AM) microwave signal with carrier frequencies between 16 and \SI{20}{\giga\hertz}. The antenna radiates circularly-polarized microwave fields along the \jso{axis of the helix}. The AM signal is imprinted onto the probe light field \jso{via the atoms} and is retrieved from the light field at the base station using a photo detector and a spectrum analyzer. 
\begin{figure}
\includegraphics[width=0.49\textwidth]{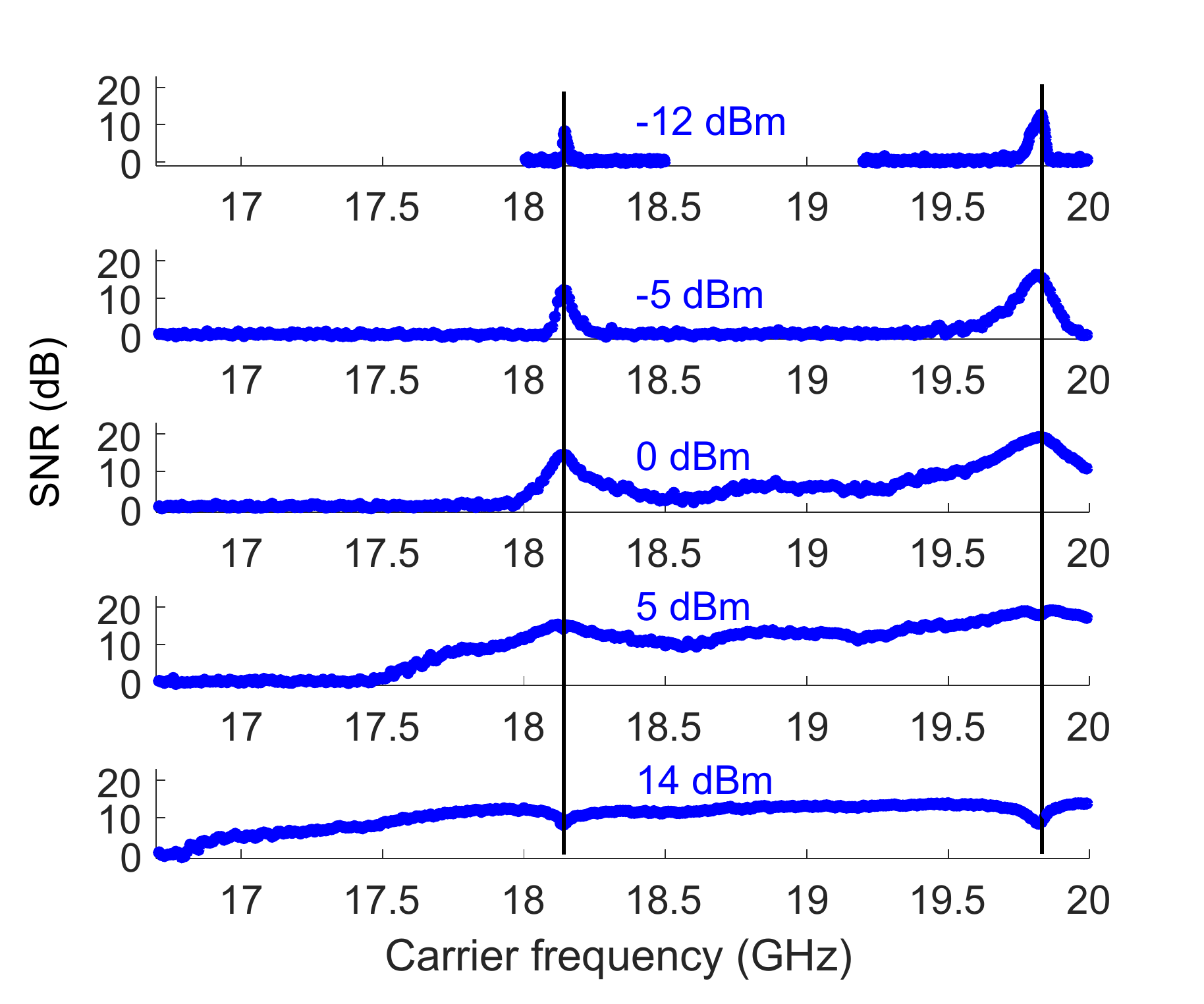}
\caption{\label{fig:scanRF} Observed SNR for a \SI{60}{\kilo\hertz} amplitude modulation, as dependent on the carrier frequencies and for different powers of the carrier field, with the sensing unit situated \SI{10}{\metre} from the base station. For the detection of the AM signal a spectrum analyzer with a \jso{resolution bandwidth of \SI{1}{\kilo\hertz} is used. The SNR is determined as the difference of the signal measured at \SI{60}{\kilo\hertz} if the AM modulation enabled or disabled}. The probe and coupling lasers are frequency stabilized to the resonant two-photon EIT transition \jso{with $\Delta_\text{p} = \Delta_\text{c} = 0$ (see Fig.~\ref{fig:setup}).}}
\end{figure}

\jso{In the following, we present measurements of our transducer setup for two settings. Figure~\ref{fig:scanRF} presents a measurement conducted in our laboratory for a distance of $\sim\SI{10}{\metre}$ between the optical table hosting the base unit and the portable sensing unit. By contrast, Fig.~3 shows field measurements in an out-of-lab setting for which the base and sensing unit were both portable (see Fig.~\ref{fig:trolley}) and separations of up to 30~m could be explored. We investigate the performance of our sensing unit for a microwave field with a carrier frequency ranging from 16~GHz to 20~GHz. While transitions between high-lying \mbox{Rydberg} states generally possess larger electric dipole matrix elements and have shown to be more sensitive for AM microwave electrometry~\cite{Cai2023} we restrict our demonstration to be utilzing the $49$D$_{5/2}$ Rydberg state (cf. Fig.~\ref{fig:setup}) due to the limited optical power of the \SI{\sim 10}{\milli\watt} coupling beam. For this choice a coupling Rabi frequency of $\Omega_\text{c} \sim 2\pi \times \SI{0.1}{\mega\hertz}$ is achieved. In absence of the RF field,  we measure an EIT linewidth of about 10\,MHz, when scanning the frequency of the coupling laser.}

\jso{Figure~\ref{fig:scanRF} shows the measured signal-to-noise ratio (SNR) as a function of the RF carrier frequency for five different RF carrier fields strengths. In these measurements, the two lasers were stabilized to the resonant two-photon EIT signal ($\Delta_\text{p} = \Delta_\text{c} = 0$) and a 60~kHz sinusoidal amplitude modulation with a modulation depth of \SI{95}{\percent} was applied to the RF field.}
For the lowest RF power (\SI{-12}{\dBm}, upper trace), we observe two distinct resonances which belong to the \jso{$49\text{D}_{5/2} \leftrightarrow 48\text{F}_{7/2}$ 
transition at \SI{19.84}{\giga\hertz},} and the $49\text{D}_{5/2} \leftrightarrow 50\text{P}_{3/2}$ transition at \SI{18.14}{\giga\hertz}. Increasing the RF power results in a  gradual widening of the resonances from initially a few tens of MHz to hundreds of MHz. \jso{Eventually}, the frequency-selective sensor turns into a broadband sensor for large RF carrier powers, as the EIT signal starts to shift due to the AC Stark effect \cite{Hu2022}. 
 \jso{A discussion of the sensing performance for off-resonant RF and optical fields is given in the supplementary material.} 
For the highest RF power (\SI{14}{\dBm}, lower trace in Fig.~\ref{fig:scanRF}) we additionally observe a dip at the resonance frequencies. This reduction in \jso{SNR} is due to the limited depth of our modulation, capped by our instrument to a maximum of \SI{95}{\percent}: the signal strength is given by the contrast between the probe transmission during AT splitting at modulation peaks and the restored EIT transmission at modulation troughs, with the latter degraded by the residual RF power.\\ 
\begin{figure}
\includegraphics[width=0.46\textwidth]{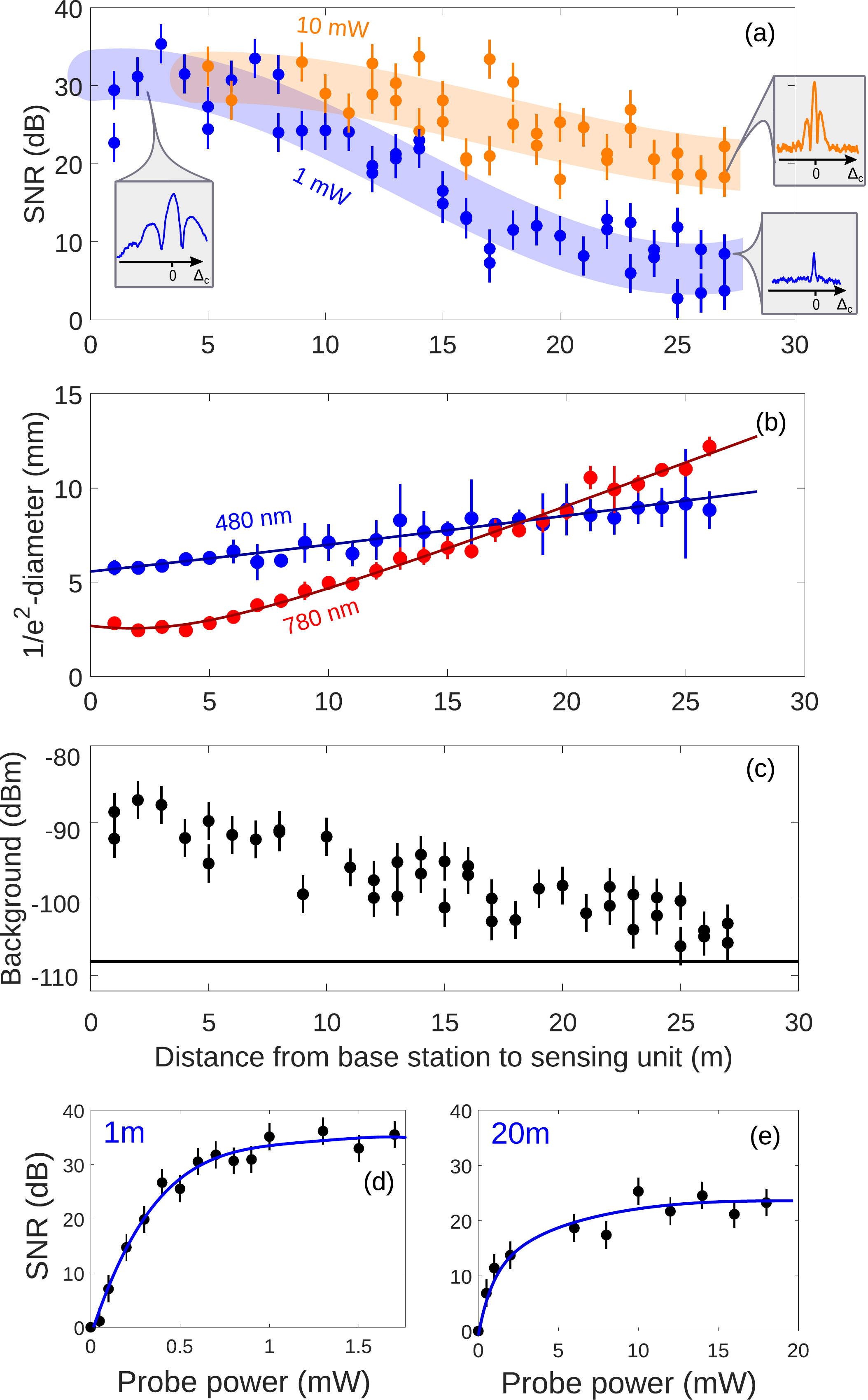}
\caption{\label{fig:SNA}\jso{(a) Peak SNR versus distance from the base station to the sensing unit for two probe coupling powers and a RF carrier power of \SI{12}{\dBm} at \SI{19.84}{\giga\hertz}. The probe power values refer to the power at the base station. The inset shows the signals detected with a spectrum analyzer in zero-span mode at \SI{60}{\kilo\hertz} while scanning the coupling laser frequency. The semi-transparent lines are a guide to the eye. (b) $1/e^2$ beam diameters of the probe and coupling beam as a function of distance to the base station. The black lines are fits for the propagation of a Gaussian beam.  (c) Reflected signal power measured at the base station for different distance of the sensing unit and for a resonant probe laser field of \SI{1}{\milli\watt}. The horizontal line shows the noise background if the probe laser is turned off.} (d) and (e) show the maximum SNR as function of probe power for the sensing unit positioned at \SI{1}{\metre} and \SI{20}{\metre} from the base station. The blue lines are a guide to the eye. }
\end{figure}
\jso{Figure.~\ref{fig:SNA} characterizes the performance of our fully portable setup (see Fig.~\ref{fig:trolley}). We transmit a \SI{12}{\dBm} RF carrier fixed at \SI{19.84}{\giga\hertz} with amplitude modulation at \SI{60}{\kilo\hertz} near the sensing unit, and we extract the signature of the modulation from the probe beam using a spectrum analyzer. While the probe laser is stabilized ($5\text{S}_{1/2} \leftrightarrow 5\text{P}_{3/2}$), the coupling laser is scanned over the 49D$_{5/2}$ transition. The maximum SNR at $\Delta_\text{c} = 0$ is determined from these scans, with examples inset in Fig.~\ref{fig:SNA}(a). If the probe power is kept to \SI{1}{\milli\watt}, we observe that the SNR drops substantially as the distance between the base station and sensing units exceeds $\SI{10}{\metre}$ (blue points in Fig.~\ref{fig:SNA}(a)). However, for these larger separations the signal can be recovered by increasing the probe power (orange points). The reduction of SNR over distance is primarily caused by the change in the beam diameters, affecting the ratio $\Omega_\text{c}/\Omega_\text{p}$ inside the vapour cell, where $\Omega \propto \sqrt{P}/d$ for a beam of power $P$ and diameter $d$. In our demonstration, the coupling beam diameter is initially twice as large as the strongly-divergent probe beam, but for distances of $>\SI{18}{\metre}$ the probe diameter exceeds the coupling beam diameter, see Fig.~\ref{fig:SNA}(b). As the distance increases, the probe divergence also results in power losses at the 1-inch detection optics on its return to the base station. This presents as a reduction in the measured background signal (noise level) if the transmitted probe power is kept constant, see Fig.~\ref{fig:SNA}(c). In Fig.~\ref{fig:SNA}(d) and (e) we show the SNR as a function of probe power for distances of \SI{1}{\metre} and \SI{20}{\metre}. To maintain $\Omega_\text{p}$ to the initial value at \SI{1}{\metre}, at \SI{20}{\metre} the probe power has to be increased by a factor of $(d_\text{20m}/d_\text{1m})^2\approx 12$. Assuming little change in $\Omega_\text{c}$ of the less-divergent coupling beam, this agrees well with the onset of saturation in the two measurements at about \SI{1}{\milli\watt} and \SI{10}{\milli\watt} respectively. Nevertheless, at \SI{20}{\metre} the achievable SNR is lower than at \SI{1}{\metre} which we attribute to an increase in the detection losses of the probe beam. Moreover, at \SI{20}{\metre} all beams are partly overlapping in the cell which results in co-linear propagating probe and coupling beams. Since we adapt the beam alignments for each distance, the each measurement in Fig.~\ref{fig:SNA}(a) involves a trade off power loss and beam overlap to get the maximum SNR achievable within a limited adjustment range.}

\jso{In general, the SNR in our demonstration is limited by the small coupling Rabi frequencies of $<2\pi\times$\SI{0.1}{\mega\hertz}. To obtain higher SNRs and a longer reach for the transducer, higher coupling powers and light fields with larger Rayleigh length $z_\text{R}$ should be used. In respect to this, an ``inverted'' EIT system (5S - 6P - nS/nD) for which the probe and coupling beam have wavelengths of $\lambda_\text{p} = $\SI{420}{\nano\metre} and $\lambda_\text{c} = $\SI{1015}{\nano\metre} respectively, will be beneficial, as $z_\text{R} \propto 1/\lambda$ and the probe beam has to travel twice the distance of the coupling beam.} In our setup, the divergence of the probe beam ($z_\text{R} \sim\SI{6}{\metre} $) will limit the free-space link to a distance of $\sim\SI{60}{\metre}$, beyond which the beam diameter will exceed the width of the cylindrical vapor cell and the 1-inch corner-cube prism reflector. \mc{Approaching this point, the SNR will decrease due to the overlap of each beam with its counter-propagating pair.} \jso{We note that a lens could be added to the setup to focus the beams into the vapor cell. This would increase the coupling intensity while reducing the risk of overlaps between the counter-propagating pairs. }

Having demonstrated the functionality of our receiver \jso{for distances of up to $\sim \SI{30}{\metre}$}, we now discuss advantages and limitations of our \jso{stand-alone transducer} system. \jso{The most} important characteristic of our portable probe is that it contains no electronic circuitry and does not require any metal parts. This reduces scattering of the RF field of interest which can limit the accuracy of measurements \cite{ Holloway2022}, for example in an anechoic chamber setting. In comparison to a fiber coupled approach \cite{Holloway2017,Simons2018,Mao2023}, our portable probe is not bound to any optical fibers and can more easily be transported, though this comes at the cost of alignment adjustments each time it is moved. Further to this, our setup benefits from a greater design flexibility. The alignment of the optical beams solely relies on the position of the corner-cube prism reflector, making it possible to use different vapor cell designs without having to modify the alignment of the optical beams. For example, a small vapor cell could be used if the spatial information of the field is desired, while a longer cell offers a large atomic volume and \jso{higher} SNR.  Although our probing scheme benefits from flexibility, an unobstructed beam pathway between the base station and \jso{sensing unit} is required. If a line of sight connection between the two stations is not possible, mirrors can be added to direct the beams to the sensing area. \jso{We note that some Rydberg-based techniques, as for example the superheterodyne technique \cite{Jing2020}, require additional RF fields to be applied to the Rydberg atom-based sensor. While it is in general possible to add a transmitting antenna and RF source to our sensing unit, this would limit its portability.}

\begin{figure}
\includegraphics[width=0.49\textwidth]{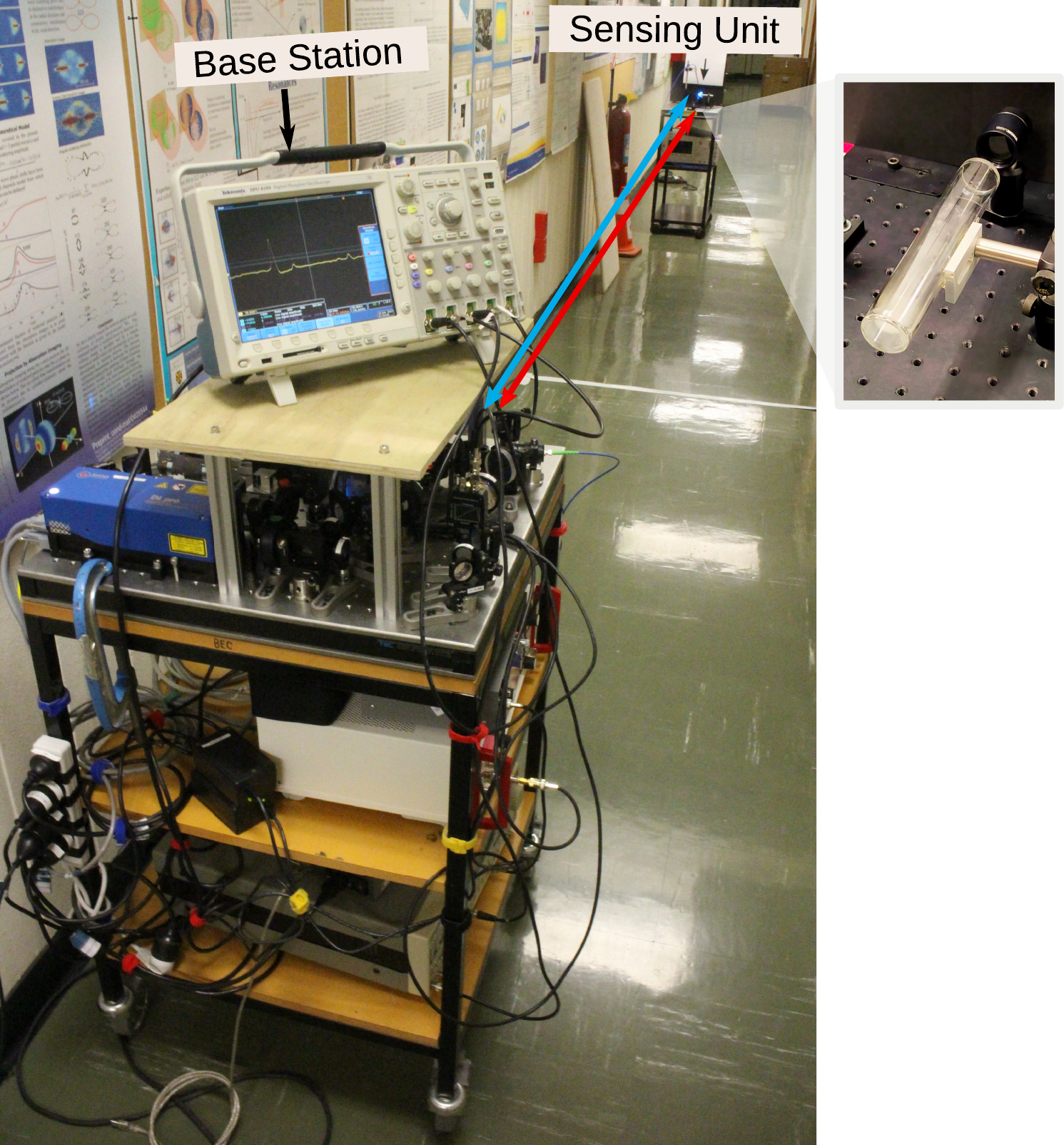}
\caption{\label{fig:trolley} Free-space-coupled setup outside our laboratory. The base station and the \jso{sensing unit} are mounted on two movable trolleys, and were placed at separations of up to \SI{30}{\metre} (pictured here at \SI{12}{\metre}). The optical board used for the base station has a footprint of \SI{1}{\metre} x \SI{1}{\metre}. \jso{The photo on the right shows the sensing} unit: a rubidium vapor cell, and a retroreflector.
}%
\end{figure}

Compared to a fiber-coupled approach, the total transmission efficiency of the probe and coupling beams can \jso{potentially} be much higher for our free-space-coupled configuration. The free-space link makes it possible to have large coupling powers at the vapor cell as only scattering in the atmosphere has to be considered. We do not measure any attenuation of the coupling and probe light fields between the base station and the vapor cell placed $\sim\SI{30}{\metre}$ apart from one another. By contrast, Ref.~\onlinecite{Simons2018} reports a transmission efficiency of the coupling beam into the vapor cell of only \SI{34}{\percent} for a coupling beam at \SI{511}{\nano\metre}. A fiber-coupled approach therefore demands higher coupling powers to overcome these losses, which might further result in nonlinear effects in the optical fibers. Of \jso{high} significance is the probe field, which carries the information and has to be guided back to the photodetector. In Ref.~\onlinecite{Simons2018}, the transmission efficiency of the probe through the system (input fiber, vapor cell, and output fiber) was only \SI{17}{\percent}. This was improved to \SI{40.4}{\percent} in Ref.~\onlinecite{Mao2023} by some changes in the design. For our free-link system, ignoring the negligible losses due to atmospheric scattering, only the transmission through the vapor cell (encountering four layers of glass over the two transits of the cell) and the reflection at the corner-cube prism reflector impact the transmitted probe power. With anti-reflection coated glass surfaces this should enable one to detect $>90\%$ of the initial probe power at the photodetector (aside from atomic absorption within the vapor cell). In our proof-of-principle demonstration several factors contributed to \jso{a lower transmission efficiency of \SI{\sim 55}{\percent}}. These include uncoated glass surfaces, adsorption of rubidium atoms on the cell surfaces, and collection optics which are smaller than the expanding probe beam upon its return. Although limiting the SNR in our demonstration \jso{(see Fig.~\ref{fig:SNA})}, none of these issues poses a fundamental limit. \jso{In contrast to the fiber-coupled approach, our free-space link allows for large beam diameters in the vapour cells. This is of advantage as more atoms can be addressed simultaneously (beneficial for the SNR), and the impact of transit-time broadening $\Gamma_\text{t}$ is strongly reduced. Assuming $\Gamma_\text{t} \approx \sqrt{2}v/d$, where $v$ is the mean velocity of the atoms~\cite{KumarGangwar2016}, in our room-temperature vapor cell $\Gamma_\text{t}$\SI{<100}{\kilo\hertz}.} This is much less than the minimum \jso{attainable} EIT linewidth due to residual Doppler shifts for a counter propagating probe and coupling beam \cite{Kuebler2018} $|k_\text{p} -k_\text{c}|/k_\text{p}\cdot\Gamma_\text{e} \approx \, $\SI{3.75}{\mega\hertz}, where $k_\text{p}$ and $k_\text{c}$ are the wave \jso{numbers} of the probe and coupling field, and $\Gamma_\text{e}$ is the lifetime of the first excited state and a scan of the probe field is assumed.

\mc{This work demonstrates a simple technique for a remote Rydberg-atom-based RF sensor which we verified for distances $\SI{30}{\metre}$ between the passive sensing unit, and the base station--- currently limited by the length of our hallway. We remark that our work shares challenges with remotely-interrogated atomic magnetometers\cite{Patton2012,Zhang2021} which operate in a similar fashion to our Rydberg microwave detector. A free-space optical setup, as discussed in this work, is applicable to these systems, though a polarization-preserving retroreflector might be needed \cite{Patton2012}.}

\section*{Supplementary Material}
See the supplementary material for a characterisation of the sensing performance of our sensing unit in a parameter space spanned by RF and optical coupling frequencies.

\begin{acknowledgments}
We thank Matthew Cloutman and Samyajit Gayen for technical assistance.
This work was supported by the Marsden Fund of New Zealand (Contracts No. UOO1923 and UOO1729) and by MBIE (Contract No. UOOX1915).
\end{acknowledgments}
\section*{Data Availability Statement}
\so{The data that support the findings of this study are available from the corresponding author upon reasonable request.}

\section*{References}
%

\begin{figure*}
\centering
\includegraphics[page=1, scale=0.99,trim={2cm 0 1cm 0},clip]{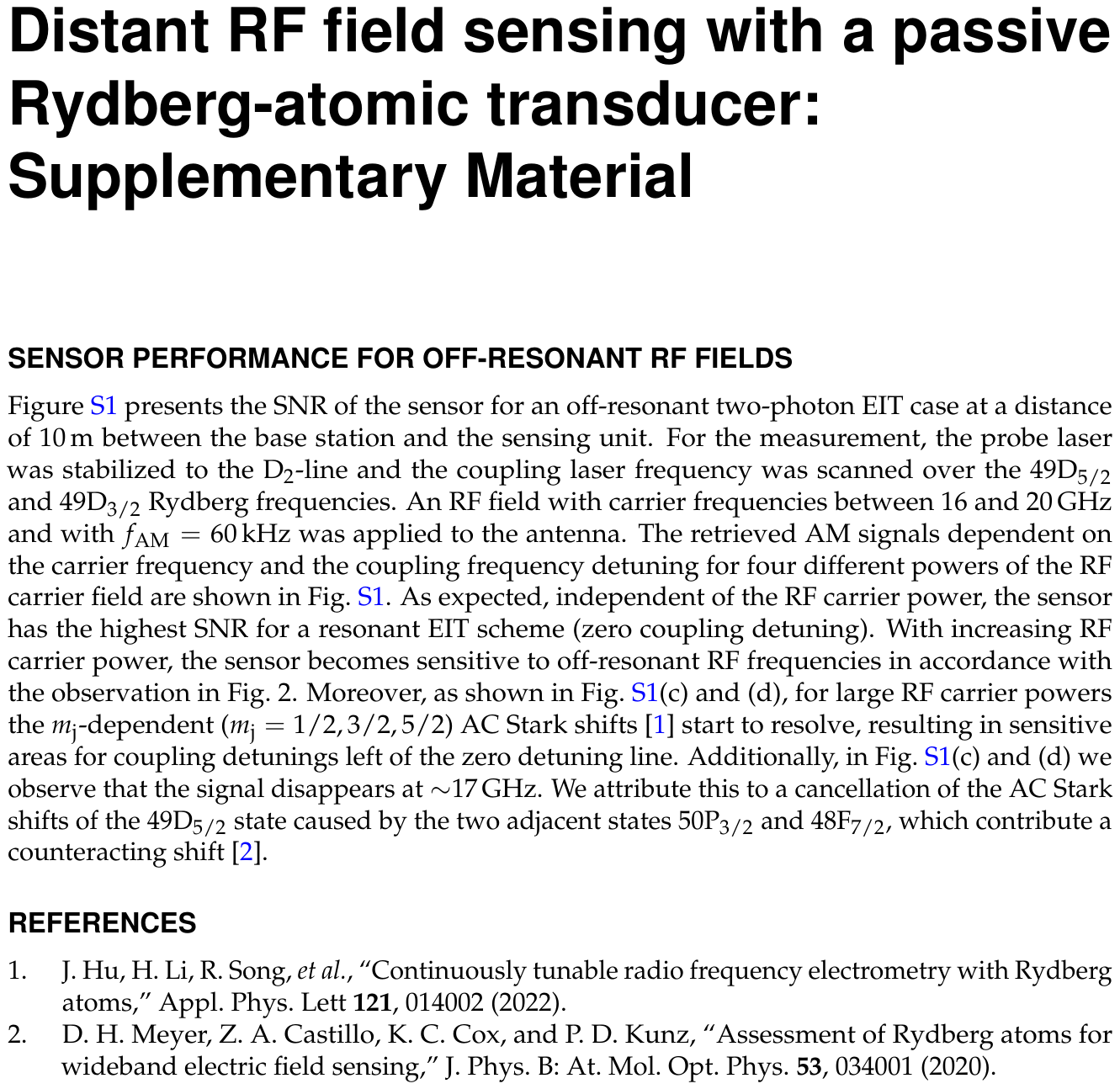}
\end{figure*}

\begin{figure*}
\centering
\includegraphics[page=2, scale=0.99,trim={2cm 0 1cm 0},clip]{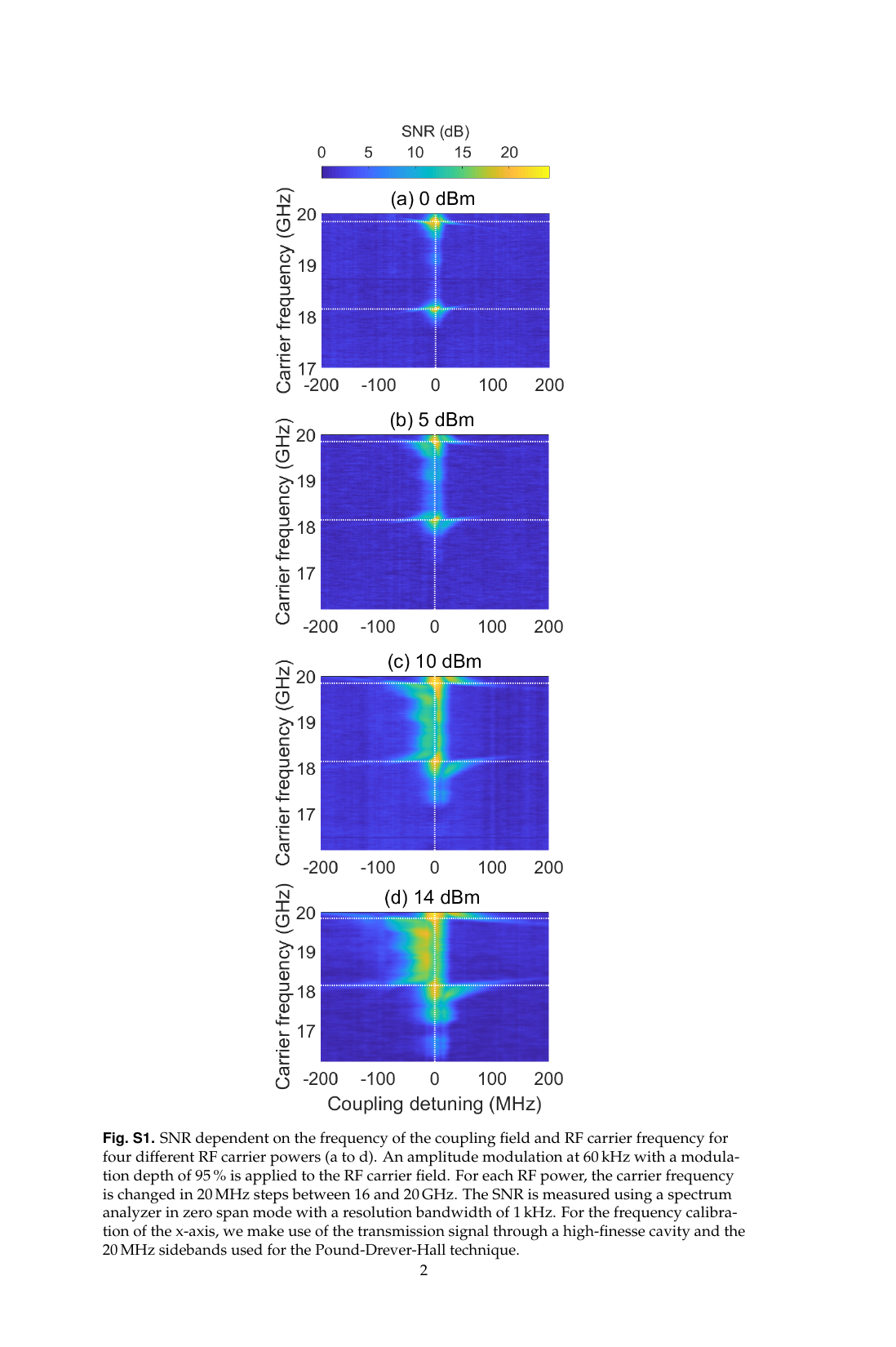}
\end{figure*}
\end{document}